\documentstyle[preprint,aps]{revtex}

\begin{document}

\draft
\title
{Fractons and Luttinger liquids}

\author
{Wellington da Cruz\footnote{E-mail: wdacruz@exatas.uel.br}}

\address
{Departamento de F\'{\i}sica,\\
 Universidade Estadual de Londrina, Caixa Postal 6001,\\
Cep 86051-970 Londrina, PR, Brazil\\}
 
\date{\today}

\maketitle

\begin{abstract}

We consider the concept of fractons as particles or quasiparticles 
which obey a specific fractal statistics in connection with a 
one-dimensional Luttinger liquid theory. We obtain a dual 
statistics parameter ${\tilde{\nu}}=\nu+1$ which 
is identified with the controlling parameter $e^{-2\varphi}$ 
of the Luttinger 
model. In this way, a bosonic system  
characterized by a fractal index $i_{f}[h]=i_{f}[2]=1$ is  
considered as a conformal field theory with central charge 
$c[\nu=0]=1=i_{f}[2]$ with a compactified 
radius $R=\frac{1}{\sqrt{\tilde{\nu}}}=1$. 
Thus, we have a mapping of a bosonic theory to a fermionic 
one and vice-versa, i.e. the duality symmetry ${{\tilde{h}}
=3-h}$ of the universal class $h$ of fractons defined in the interval 
$1$$\;$$ < $$\;$$h$$\;$$ <$$\;$$ 2$ is satisfied.

\end{abstract}


We have introduced in\cite{R1} the concept of universal classes of particles or 
quasiparticles, labeled by a fractal parameter ( or Hausdorff dimension ) 
$h$, defined in the interval 
$1$$\;$$ < $$\;$$h$$\;$$ <$$\;$$ 2$. These particles termed 
{\it fractons}\footnote{ This word was used by Alexander and Orbach 
in another context\cite{R8}.} 
obey {\bf fractal statistics} 
( the distribution functions are {\it fractal functions} )

\begin{eqnarray}
\label{e.h} 
n[h]=\frac{1}{{\cal{Y}}[\xi]-h},
\end{eqnarray}

\noindent where the function ${\cal{Y}}[\xi]$ satisfies 
the equation 

\begin{eqnarray}
\label{e.46} 
\xi=\left\{{\cal{Y}}[\xi]-1\right\}^{h-1}
\left\{{\cal{Y}}[\xi]-2\right\}^{2-h},
\end{eqnarray}
 
\noindent and $\xi=\exp\left\{(\epsilon-\mu)/KT\right\}$, 
has the usual definition. The particles are collected 
in each class taking into account the {\it fractal spectrum}

\begin{eqnarray}
\label{e.7}
&&h-1=1-\nu,\;\;\;\; 0 < \nu < 1;\;\;\;\;\;\;\;\;
 h-1=\nu-1,\;
\;\;\;\;\;\; 1 <\nu < 2;\\
&&etc.\nonumber
\end{eqnarray}

\noindent and the spin-statistics relation $\nu=2s$.

Fractons are charge-flux systems which live in two dimensional multiply 
connected space and carry rational or irrational quantum numbers, as charge and spin. 
The fractal parameter $h$ is associated to the path of 
the quantum-mechanical particle and so, we have a {\it quantum-geometrical} 
description of the statistical laws of Nature. This means that the 
fractal characteristic of the quantum path, which reflects the 
Heisenberg uncertainty principle, is embodied in the fractal 
statistics expression. Thus, our formulation generalizes 
{\it in a natural way} the 
bosonic and fermionic distributions. Besides this, 
we verify that the classes $h$ of particles have a 
{\it duality symmetry} defined by ${\tilde{h}}=3-h$, such that fermions ($h=1$) 
and bosons ($h=2$) are dual objects. As a result, we extract a {\it 
fractal supersymmetry} explicited by the {\bf theorem}: 
{\it If the particle with spin 
$s$ is within the class $h$, then its dual $s+\frac{1}{2}$ is into the class 
${\tilde{h}}$}.

Now in\cite{R2} we have obtained a relation between fractons and 
conformal field theory(CFT)-quasiparticles ( edge excitations ), 
through the concept of {\it fractal index} associated with 
the classes $h$ 

\begin{equation}
\label{e.1}
i_{f}[h]=\frac{6}{\pi^2}\int_{\infty(T=0)}^{1(T=\infty)}
\frac{d\xi}{\xi}\ln\left\{\Theta[\cal{Y}(\xi)]\right\}
\end{equation}

\noindent where

\begin{eqnarray}
\Theta[{\cal{Y}}]=
\frac{{\cal{Y}}[\xi]-2}{{\cal{Y}}[\xi]-1}
\end{eqnarray}

\noindent is the single-particle partition function of the 
universal class $h$ and hence

\begin{eqnarray}
\label{e.h} 
n[h]&=&\xi\frac{\partial}{\partial{\xi}}\ln\Theta[{\cal{Y}}].
\end{eqnarray}

\noindent We have verified a connection between 
the central charge $c[\nu]$(a dimensionless number which 
characterizes conformal field theories in two dimensions) and 
the universal classes $h$ 
of particles. For that we considered the particle-hole duality 
$\nu\longleftrightarrow\frac{1}{\nu}$ for integer-value 
$\nu$ of the statistical parameter which some systems 
of the condensed matter satisfy, for example, the 
Calogero-Sutherland model\cite{R3} and others\cite{R4}. 
The central charge for $\nu$ {\it even} is defined by

\begin{eqnarray}
\label{e.11}
c[\nu]=i_{f}[h,\nu]-i_{f}\left[h,\frac{1}{\nu}\right]
\end{eqnarray}

\noindent and for $\nu$ {\it odd} is defined by

\begin{eqnarray}
\label{e.12}
c[\nu]=2\times i_{f}[h,\nu]-i_{f}\left[h,\frac{1}{\nu}\right],
\end{eqnarray}

\noindent where $i_{f}[h,\nu]$ means the fractal 
index of the universal class $h$ which contains the particles 
with distinct spin values which obey a specific 
fractal statistics.

In this short note we show the connection between fractons 
and Luttinger liquids, taking into account the 
parameters which characterize these models. They are one-dimensional 
interacting many-body systems, which at low-energy or 
low-temperature are characterized by a controlling parameter, 
$e^{-2\varphi}$\cite{R5}.

We also have obtained a Fermi velocity given by 
 $v_{F}=\frac{v_{N}}{{\tilde{\nu}}}$, where 
 ${\tilde{\nu}}=\nu+1$ ($\nu\geq 0$) is a dual statistical 
 parameter and $v_{N}=\frac{2\hbar}{mL}c[\nu]$ is an excitation 
 related to the change in the particle number which 
 depends on the central charge\cite{R2}. The quantities $\hbar$ 
is the Planck constant, $m$ is the mass of particle and $L$ is the 
length of the edge. Another excitation related to the current 
is defined as 
$v_{J}=\frac{v_{F}}{\tilde{\nu}}$. Haldane\cite{R5} has just 
considered a relation 
between these two types of excitation for characterize the 
Luttinger liquid 
$v_{F}={\sqrt{v_{N}v_{J}}}$. Thus, our definition satisfies 
this condition 
and on one way, we identify the controlling parameter 
with the dual statistical parameter ${\tilde{\nu}}=e^{-2\varphi}$, 
such that $v_{N}=v_{F} e^{-2\varphi}$ and $v_{J}=v_{F} e^{2\varphi}$.

Now, we observe that ${\tilde{\nu}=\nu+1}$($\nu\geq 0$), satisfies 
the duality symmetry, i.e. we can map a fermionic theory 
to a bosonic one and vice-versa. We also know that the fractal 
index $i_{f}[h]$ coincides with the central charge $c[\nu]$, for 
$0 \leq\nu\leq 1$ and we have distinguished two concepts 
of central charge, one related to the universal classes $h$ 
and the other related to the particles which belong 
to these classes\cite{R2}. Thus, the fractal index for 
$\nu=0$( boson, $\nu$ even ) is $i_{f}[h,\nu]=i_{f}[2,0]=1$ 
and for the dual statistical parameter ${\tilde{\nu}}=1$
( fermion, ${\tilde{\nu}}$ odd ) is  $i_{f}[{\tilde{h}},
{\tilde{\nu}}]=i_{f}[1,1]
=\frac{1}{2}$. In this way, 
we define for a bosonic theory $i_{f}[2]=1$ identified with 
a conformal field theory with central charge $c[\nu=0]=1$, a 
compactified radius $R$ determined by the dual 
statistical parameter $R^2=\frac{1}{{\tilde{\nu}}}=e^{2\varphi}$ and 
$\varphi=\ln{R}$. The radius $R$ expresses an angular 
invariance of the theory.

Finally, we would like {\it to emphasize} that our results were 
obtained independently of any bosonization idea as 
discussed in\cite{R6}. In contrast, we have used a {\bf symmetry principle}, 
i.e. the concept of {\bf duality symmetry} between the universal 
classes $h$ of particles. On the other hand, our definition 
of {\bf fractal statistics} differs of the notion 
of {\it exclusion statistics}\cite{R7}, given that in our approach 
we consider {\it ab initio} the spin-statistics relation 
$\nu=2s$ for particles with rational or irrational  spin quantum 
number $s$. This claim is supported by the {\it fractal spectrum} 
which is a {\it mirror symmetry} in our formulation. Another consequence 
is the {\it fractal supersymmetry} which is realized, for example, in the 
context of fractional quantum Hall effect\cite{R1}.

\end{document}